\documentclass[12pt]{iopart}
\eqnobysec
\usepackage{graphicx}
\usepackage{color}
\usepackage{subcaption}
\usepackage{bm}

\font\amsb=msbm10
\def\hbar{\mbox{\amsb\char'175}}

\newcommand{\be}{\begin{equation}}
\newcommand{\ee}{\end{equation}}

\newcommand{\vecp}{{\mathbf p}}
\newcommand{\vecq}{{\mathbf q}}
\newcommand{\vecI}{{\mathbf I}}
\newcommand{\Vxi}{{\boldsymbol \xi}}
\newcommand{\x}{{\mathbf x}}

\newcommand{\X}{{\mathbf X}}

\newcommand{\y}{{\mathbf y}}

\newcommand{\J}{{\mathbf J}}

\newcommand{\C}{{\mathbf C}}

\newcommand{\vP}{{\mathbf P}}
\newcommand{\vQ}{{\mathbf Q}}

\newcommand{\opA}{{\widehat{A}}}

\newcommand{\opR}{{\widehat{R}}}
\newcommand{\opT}{{\widehat{T}}}

\newcommand{\opH}{\widehat{H}}
\newcommand{\opU}{\widehat{U}}
\newcommand{\opI}{\widehat{I}}

\newcommand{\M}{{\mathbf M}}

\newcommand{\der}{\partial}

\begin{document}

\title{On the classical geometry of chaotic Green functions and Wigner functions}

\author {A. M. Ozorio de Almeida\footnote{ozorio@cbpf.br}}
\address{Centro Brasileiro de Pesquisas Fisicas, 
Rua Xavier Sigaud 150, 22290-180, 
Rio de Janeiro, R.J., Brazil.}

\begin{abstract}

Semiclassical approximations for various representations of a quantum state are constructed on a single (Lagrangian) surface in phase space, but it is not available for chaotic systems.
An analogous {\it evolution surface} underlies semiclassical representations of the evolution operator,
albeit in a doubled phase space. It is here shown that, corresponding to 
the Fourier transform on a unitary operator, represented as a Green function or spectral Wigner function, 
a Legendre transform generates a {\it resolvent surface} as the classical basis 
for semiclassical representations of the resolvent operator in double phase space,
independently of the integrable or chaotic nature of the system. This surface coincides with 
derivatives of action functions (or generating functions) depending on the choice of appropriate coordinates
and its growth departs from the energy shell following trajectories in double phase space.
In an initial study of the resolvent surface based on its caustics, 
its complex nature is revealed to be  analogous to a multidimensional sponge.

Resummation of the trace of the resolvent in terms of linear combinations of periodic orbits, 
known as pseudo orbits or composite orbits, provides a cutoff to the semiclassical sum at the Heisenberg time.
It is here shown that the corresponding actions for higher times can be approximately included within true {\it secondary periodic orbits},
in which multiple windings of short periodic orbits are joined by heteroclinic orbits into larger circuits.

\end{abstract}

\maketitle

\section{Introduction} 

The semiclassical (SC) theory for unitary evolution of the states of a closed quantum system has a long continuing history, 
starting with Van Vleck in 1928 and proceeding to contemporary computations such as Lando et al. 2024 \cite{VVleck28,Maslov,livro,Gutzwiller,Haake,Lando}. 
It is constructed on the corresponding evolution
of privileged surfaces in the classical phase space. For instance, in the case of a
position state $|\vecq\rangle$ with $\vecq= (q_1,...,q_L)$,  the initial state $|\vecq_-\rangle$ 
corresponds to the $L$-dimensional plane $\vecq=\vecq_-$ in the $2L$-dimensional phase space
with points $\x=(\vecq, \vecp)$. Driven by a Hamiltonian $\widehat H$, the initial state evolves
in time under the action of the evolution operator
\be
\opU_t= \exp{\left(-\frac{it}{\hbar}\opH \right)},
\label{opevol}
\ee
corresponding to a continuous group of canonical transformations $\C_t$, that is, $\C_t:\x_-\mapsto \x_+$,  
generated by the classical Hamiltonian $H(\x)$ according to Hamilton's equations.

The evolved classical surface $\x_+= {\tilde \x}(\vecq_-,\vecp_-,t)$, corresponding to the evolved position state $|\vecq_-(t)\rangle$, 
will not generally be a plane. However, even if the initial surface has a more complex topology, it will preserve the Lagrangian property:
the action for any reducible closed loop $\gamma$ on the surface is zero, i.e.
\be
S_\gamma =\oint_{\gamma} \vecp \cdot \rmd \vecq = 0 ~. 
\label{Lag}
\ee
Furthermore, the surface is explicitly defined by the derivatives of a scalar generating function $S(\vecq_+,t)$:
\be
\vecp_+ = \tilde{\vecp}(\vecq_-,\vecp_-,t)=\frac{\der S(\vecq_+,t)}{\der \vecq_+}.
\ee

Other examples are the eigenstates of integrable Hamiltonians, which, according to Arnold's theorem, 
correspond to Lagrangian $L$-dimensional tori \cite{Arnold78},
so that their generating function and hence their projection (on the position plane $(\vecp=0)$ or any other Lagrangian plane) is many valued, with the sheets joined along caustics (see e. g. \cite{livro}). On the other hand, no Lagrangian surface 
can be paired to the eigenstate of a classically chaotic system in the $2L$-dimensional phase space.

For this reason, the present semiclassical theory reaches out into double phase space 
\footnote{Also termed the secondary phase space \cite{OsKon}.},  
whose elements are all the ordered pairs of phase space points and hence encompasses
all possible classical transitions. For example, a uniform translation of the ordinary phase space
is represented by a Lagrangian plane in double phase space, 
which is transverse to the double phase space plane that defines a canonical reflection through a phase space point. 
The points on these sets of planes can be used as conjugate coordinates for double phase space, which correspond,
respectively, to the chord and the centre (Weyl) representations (reviewed in appendix A). In contrast, 
the planes $(\vecq=\vecq_+, \vecq=\vecq_-)$ and $(\vecp=\vecp_+, -\vecq=\vecq_-)$
are also Lagrangian coordinate planes, but they do not represent canonical transformations.

First, it is necessary to review the role of a general Lagrangian {\it evolution surface} in the double $4L$-dimensional phase space, 
which contains all pairs of points connected in a given time by the canonical transformation generated by the Hamiltonian. 
It is the backbone of SC approximations of the unitary evolution operator. Then this paper goes on to present the Lagrangian {\it resolvent surface} in a doubled phase space, composed of all pairs of points connected by trajectory segments in a given energy shell $H(\x)=E$.
This, in its turn, corresponds to the energy dependent {\it resolvent operator}
\be
\widehat{U}_E = [\widehat{H}-E \hat{I}]^{-1}= \sum_n \frac{|n\rangle \langle n|}{E_n - E} 
= \frac{1}{2\pi\hbar} \int_{-\infty}^{\infty} {\rm d}t  ~\exp \left[\frac{i}{\hbar} Et \right] ~ \widehat{U}_t ~.
\label{resolvent1}
\ee 
The poles of the resolvent operator lie on the eigenergies of $\opH$, that is,
\be 
{\rm tr} ~ {\widehat U}_E = \sum_n \frac{1}{E_n - E}
\ee
and the residues are the projectors onto its eigenstates. Thus, all we need in principle is to construct
a semiclassical approximation on this previously unexplored $2L$-dimensional Lagrangian surface, 
in analogy to the SC representations of states in terms of their $L$-dimensional Lagrangian surfaces.
Even though this is more complex than the integrable case, this surface is again obtained locally
by $2L$ derivatives of various generating functions in related coordinate systems of the double phase space.

Considering that the resolvent operator is the Fourier transform of the evolution operator, it is only natural
that the resolvent surface should arise in the SC theory through the Legendre transformation of the evolution surface.
This is implicit in the local SC derivation of the phase of the resolvent operator as an action or generating function, 
but it does not seem to have yet been brought to the fore geometrically.
Actually, the easiest grasp on the the $2L$-dimensional resolvent surface is to keep track of its growth  
from the $(2L-1)$-dimensional energy shell within the initial {\it identity plane} $H(\x_+)=H(\x_-)=E$, 
by following in time all its trajectories $\x_+= {\tilde \x}(\vecq_-,\vecp_-,t)$.  
In spite of its growing complexity, this is a single smooth surface for all time.
\footnote{The finite time cutoffs of the resolvent surface already support the SC approximation
of energy smoothed Green function and Wigner functions.}
It will be shown here that the initially simple 'cylindrical extension' of the energy shell into double phase space
eventually develops folds along each of its periodic orbits (po's). 

Ideally, all the po's of chaotic systems are isolated and hyperbolic, endowed with stable and unstable manifolds
composed of orbits, which approach the po asymptotically as $t\rightarrow \infty$ or $t\rightarrow -\infty$ respectively.
In reality, there will often be tiny islands centred on elliptic orbits, but these will not be further considered here.
The stable and unstable manifolds of the same po (or different po's) intersect along homoclinic orbits (or heteroclinic orbits),
forming a dense net of connections among the multitude of po's. It will be argued that each closed loop formed by these connections, 
joined to arbitrary numbers of windings among each of the internal po's in the given circuit, 
is very close to a bona fide {\it secondary periodic orbit}. Each one of these is again responsible for a fold in the resolvent surface
as its period is reached. Furthermore, short open orbits, which contribute to the Green function and the Wigner function,
also come arbitrarily close to these circuits with their multiple windings, which can also be included into {\it secondary open orbits}. 

In the limit of large windings for all the component po's of a secondary po, the contribution to the total action from its po windings
may be identified with the action of a {\it pseudo orbit} \cite{Ber-Keat90,Ber-Keat92} or {\it composite orbit} \cite{Bog92,Sieber07} that, 
so far, has only been a construction for the resummation of the resolvent and its trace in terms of spectral determinants. 
Thus, one can now reinterpret these collective apparently disjoint po contributions to the resolvent 
as a resummation of secondary po's that are already present in the representations of the resolvent operator and its trace.

The following section introduces the employment of double phase space in the SC approximation of unitary operators,
presenting the evolution surface. Then the Legendre transform, corresponding semiclassically to the Fourier transform
between the evolution operator and the resolvent operator, generates the resolvent surface in the next section. 
Section 4 then discusses the various representations of the resolvent operator, corresponding semiclassically to different choices 
of canonical coordinate planes in double phase space. Special attention is bestowed on the Wigner-Weyl representation (reviewed in Appendix A), 
in which the basic double phase space coordinate plane coincides with the identity plane, from which the resolvent surface grows. 
Then the simple example of a system with a single degree of freedom is examined in section 5. Leaving the special case 
of integrable systems to Appendix B, section 6 then jumps into the multiply folded resolvent surface
of a chaotic system. Finally, sections 7 and 8 present respectively the secondary periodic orbits and their 
presence in the representations of the resolvent operator and its trace.

My first presentation of the chaotic resolvent surface as a kind of sponge in double phase space was at a meeting in honour of John Hannay's sixtieth birthday in Bristol in 2011. This article for the special issue in honour of Viktor Dodonov marks its appearance in print.

\section{Review of the double phase space scenario for unitary operators}

Time dependent unitary operators that act on quantum states in Hilbert space correspond classically
to evolving canonical phase space transformations. In the case of motion generated by a 
constant Hamiltonian operator, $\opH$, the continuous group of unitary operators \eref{opevol} 
transport linearly the Hilbert space vectors $|\psi_t\rangle=\opU_t|\psi_0\rangle$. 
The various representations of the unitary operators correspond semiclassically
to different generating functions, which determine the phase of the 
corresponding quantum propagators. The Schr\"odinger equation for $\opU_t$,
e.g. in either of the position, momentum, or the Weyl-Wigner representations, are matched to 
corresponding versions of the Hamilton-Jacobi equation (see e.g.\cite{livro}). 

It might seem perverse to double the phase space of classical mechanics,
which is already a doubling of position space. Nonetheless, we are here
concerned with representing operators, commonly represented by both {\it bra} and
{\it ket} spaces, so it is not surprising that classical correspondence
generally calls for a doubled classical space. Observables are deceptively simple in 
this respect (see e.g. \cite{Alm98}, but, corresponding to unitary transformations, there arises 
an attractive and simple geometrical picture for the canonical transformations,
$\C_t:\x_-\rightarrow \x_+$, defined in the original phase space. Indeed the 
canonical property demands that all closed curves, $\gamma_-$,  be mapped
onto closed curves, $\gamma_+$, such that
\be
\oint_{\gamma_-} \vecp_- \cdot \rmd \vecq_- = \oint_{\gamma_+} \vecp_+ \cdot \rmd \vecq_+.
\label{canonical}
\ee  
Therefore, the definition of the double momentum space, $\vP=(-\vecp_-, \vecp_+)$,
and the double positions, $\vQ=(\vecq_-,\vecq_+)$, allows us to reinterpret the canonical 
condition as
\be
\S_\Gamma \equiv \oint_{\Gamma} \vP\cdot \rmd \vQ = 0,
\label{canonical2}
\ee
where $\Gamma=(\gamma_-,\gamma_+)$. These are arbitrary closed curves on 
the $(2L)$-dimensional surface defined by the one-to-one function, $\x_+=\C_t(\x_-)$,  
within the $(4L)$-dimensional {\it double phase space} $\X=(\vP,\vQ)$.
Figure 1 supplies a two-dimensional portrayal of the double coordinate system,

\begin{figure}
\centering
\includegraphics[width=0.5\linewidth]{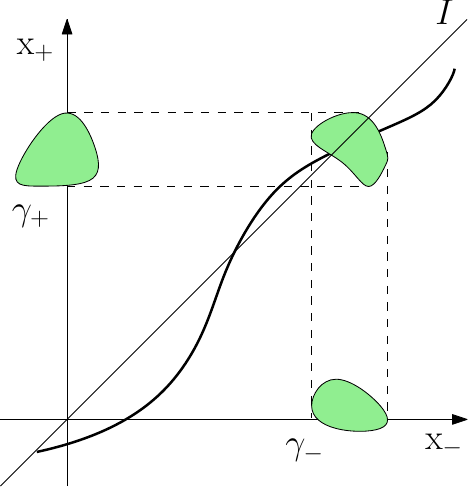}
\caption{Representation of double phase space in two dimensions, such that 
$L$-dimensional planes become straight lines. The $\x_\pm$-planes are not Lagrangian, but the diagonal line $\x_+ = \x_-$ represents the Lagrangian 
identity plane (or the centre plane). It evolves as the Lagrangian evolution surface, also sketched. Both curves $\gamma_\pm$ are projections of a curve $\Gamma$ on the evolution surface.
\label{Fig1}}
\end{figure}

The fact that the action, or symplectic area  $\S_\Gamma=0$ for any closed curve drawn 
on the surface that defines the canonical transformation $\C_t$ signifies that canonical transformations are described 
by $(2L)$-dimensional {\it Lagrangian evolution surfaces} \cite{AbrMar} in the $(4L)$-dimensional double phase space. This Lagrangian property
allows us to define locally a function,
\be
\S_t(\vQ)=\int_{\vQ_0}^\vQ \vP_t(\vQ)\cdot \rmd \vQ,
\ee
which is independent of the path followed in the $(2L)$-dimensional position coordinate plane between $\vQ_0$ and $\vQ$,
in complete analogy to the ordinary single phase space action
\be
S(\vecq)=\int_{\vecq_0}^\vecq \vecp(\vecq)\cdot \rmd \vecq ~.
\ee
In its turn, this {\it generating function} defines the given evolution
surface by the equations
\begin{eqnarray}
\frac{\der \S_t}{\der \vQ}=\vP_t(\vQ),\>\>
{\rm or}\>\>{\der \S_t\over{\der \vecq_+}} = \vecp_+,\>\>{\der \S_t\over{\der \vecq-}}=-\vecp_-,
\label{generate}
\end{eqnarray}
that determine implicitly the canonical transformation \cite{Goldstein,Arnold78}. Furthermore,
one obtains the energy of the trajectory segment as $\der \S_t/\der t=E$.
 
Though the mapping $\C_t:\x_-\rightarrow \x_+=\C_t(\x_-)$ is necessarily
univalued, no such restriction results on the function $\vP_t(\vQ)$, 
defined by the same evolution surface. (We cannot define 
generating functions using $\x_-$ or $\x_+$ as independent variables,
because neither of the planes $\x_{\pm}=0$ is Lagrangian in the double phase space.)
What is allowed and often desirable is to apply
linear canonical transformations to the double phase space,
$\X\rightarrow \X'$, which leave invariant the Lagrangian property
for any surface, including $\vP'=0$. Then we may define
a new generating function in the new variables, 
such that the invariant evolution surface is represented locally as $\vP'_t(\vQ')= \der \S'_t/ \der \vQ'$. 

All the commonly used generating functions \cite{Goldstein} are obtained
by the application of canonical $90^o$ rotations to the single phase spaces,
$\vecq_{\pm}\rightarrow \vecp_{\pm}$, separately or in combination. Obviously,
there exist unlimited other possibilities in double phase space \cite{AmHu80},
but we will here be concerned only with the special Wigner-Weyl canonical variables.
So, recalling the skew product,
\be
\x\wedge \x'=\sum_{n=1}^L (p_l q'_l - q_l p'_l)= \J\>\x \cdot \x', 
\label{squew}
\ee
which also defines the skew symplectic matrix $\J$, we construct the new canonical variables $\X \rightarrow(\vQ',\vP')=(\x,\y)$ in double phase space
\be
\vQ'=\x=\frac{\x_+ + \x_-}{2}, \>\>\vP'=\y= \J(\x_+ -\x_-)=\J{\Vxi} ~,
\label{rotation}
\ee 
where $\x$ is the {\it centre} and $\Vxi$ is the {\it chord} of the transformation. 
Instead of the previous $90^o$ rotations, this is more like a $45^o$ rotation in double phase space.
The plane $\y=0$ (or $\Vxi=0$) clearly specifies the identity transformation, $\vecI$,
which corresponds to the Lagrangian {\it identity plane} shown in Figure 1. Actually, all planes $\y=constant$ are
uniform translations, $T_\Vxi$, by the vector $-\J\y=\Vxi$, whereas each plane
defined by a constant $\x$ identifies the reflection $R_\x: \x_{-} \rightarrow \x_{+}=-\x_- +2\x$. 
Unlike the Lagrangian plane $(q_-,q_+)$, the planes $\y=0$ and $\x=0$ can be considered 
as phase spaces on their own: the space of reflection centres 
(Weyl space) and the space of translation chords.
\footnote{There are definite advantages in defining $\Vxi$ as a half chord, as in \cite{SarOA16},
but this would involve an adaptation from most of my own references.}

However, it must be remembered that these planes are Lagrangian, as far as the double 
phase space action $\S_t$ is concerned. 
Therefore, the mapping $\C_t$ defines implicitly the local evolution surface as the
function $\y(\x,t)$ in terms of the generating function $\S_t(\x)$:
\be
\y(\x,t)=\J\Vxi_t(\x)=\frac{\der \S_t}{\der\x}
\ee. 
\footnote{Alternatively, the generating function $\S_t(\y)$ can be defined, 
such that $\x(\y,t)=\der \S_t/\der\y$, corresponding to an analogously defined chord generating function \cite{Alm98}.}
It is important to point out that in this context the Lagrangian $\x$-plane registers the centres
of pairs of points $\x_\pm$, fully determined by the chord $\Vxi(\x)$ according to \eref{rotation}. 
For this reason it can be preferable to refer here to the centre plane, rather than to the identity plane.
In particular, a full return with $\Vxi=0$, that is, $\x=\x_+=\x_-$ only occurs if $\x_-$ lies on a periodic trajectory.

The evolution surface is generated continuously by the {\it Schr\"odinger} double phase space classical Hamiltonian \cite{OsKon}
\be
I\!\!H(X)=I\!\!H(\x,\y)=H(\x-\J \y/2)=H(x_+)~,
\label{Scham}
\ee
which moves $\x_+$ forward in time from the initial surface $\x_+ =\vecI(\x_-) =\x_-=\x$, while keeping $\x_-$ fixed.
\footnote{My previous work has dealt with an alternative Heisenberg double phase space Hamiltonian, corresponding
to super-operators driving the evolution of quantum operators \cite{AlmBro06,SarOA16}.}
Hence, the trajectory from the initial to the final point is $\x_+=\tilde{\x}(\x_-,t)$, 
whereas the trajectory in double phase space is ${\tilde\X}(t)= (\tilde{\x}(\x_-,t),\x_-)$.
As previously noted, the initial surface is just the Weyl coordinate surface
$\y=0$, so that, for short times, the evolution surface surface $\y(\x,t)$ will remain a one-to one function.
For longer times, this surface may fold in many ways, so that it is expressed in an action function $\S_t(\x)$ with many branches, 
which also arise in the other possible classical representations $\vP(\vQ,t)$ of the evolution surface. 
On the other hand, the fact that the evolution
surface is diffeomorphic to the identity plane guarantees that no wrappings or handles develop on the evolution surface itself.

\section{Extended double phase space}

The evolution surface contains all pairs of points in double phase space, which are joined by a trajectory segment
in the time $t$, that is, $\x_+= \tilde{\x}(\x_-,t)$. Since each segment has the constant energy $H(\x_-)=H(\x_+)=E$,
one may consider that the evolution surface is the time slice of a full $(2L+1)$-dimensional extended evolution surface
in the $2(2L+1)$-dimensional space, by supplementing double phase space with the canonical variables $t$ and $E$ \cite{Arnold78}.
The extended evolution surface is defined by the trajectory segments of all durations $t$ and all energies $E$.

It is then conceivable to choose the energy $E$ as the free variable, with the dependent time $t=t(\x_-,E)$
for each point on the extended surface. Furthermore, one may then consider its constant energy slices, even if
the resulting surface, composed of all pairs of points $\x_+=\tilde{\x}(\x_-,E)$ irrespective the time,
need not have the simple topology deduced for the evolution surface. Then, even if it has many branches,
one can locally derive the energy action as a Legendre transform of the previous action in time:
\be
\S_E(\vQ) = \S_t(\vQ) -t E
\label{enaction}
\ee
with $t(E)$ determined by  
\be
\frac{\der \S_t}{\der t}(\vQ) = E 
\ee
for each branch.
This is again a generating function, defining implicitly the local transformation on the energy shell
in the same form as \eref{generate}, that is, for $\vQ=(\vecq_-, \vecq_=)$, the derivatives of the energy action
supply the corresponding momenta:
\be
\vP(\vQ,E)= \frac{\der \S_E}{\der \vQ}(\vQ) ~.
\label{localsurf}
\ee
Likewise, for the alternative choice $\vQ'=\x$, the derivative of the corresponding generating function is the chord $\Vxi(\x,E)$ of the trajectory segment of energy $E$ centred on $\x$.  
On the other hand, the duration of the same segment determined by $(\vQ,\vP)$  or $(\vQ',\vP')$ is 
\be
\frac{\der \S_E}{\der E}(\vQ) =t ~~~ {\rm or} ~~~ \frac{\der \S_E}{\der E}(\x) =t 
\ee

The  $(2L)$-dimensional manifold, that is represented by \eref{localsurf} in each Lagrangian coordinate system, shall be referred to
as the {\it resolvent surface}, because of its quantum manifestation in the following section.   
It is important to distinguish this manifold from the mere tensor product of the pair of energy shells: $H(\x_\pm =E)$.
The latter contains all pairs of points in the ordinary shell $H(\x)=E$ and is $2(2L-1)$-dimensional. Of course, this includes
the $(2L)$-dimensional resolvent surface of pairs of points connected by a trajectory segment of energy $E$.
Indeed, given that all segments in the ordinary energy shell are included, one may follow the growing resolvent surface from
the single $(2L-1)$-dimensional energy shell on the identity plane: $H(\x_+)=H(\x_-)=E$ at $t=0$.
Then for very short times $t'$, the resolvent is a cylinder with a spherical base: 
\be
\x_+=\x_- + t'~ \dot{\x}_+(H(\x_+=\x_-=\x)=E)= ~t' ~\J \frac{\der H}{\der \x} \Big{|}_{H(\x)=E} ~. 
\label{initialcyl}
\ee
The smooth elongation in time of the continuum of trajectories, which composes it, cannot change this basic topology, 
no matter how intricately the resolvent surface folds itself.

The development of the resolvent surface from this seed will be taken up in later sections. For now, one should note that
the distinction with respect to the product of two shells becomes dramatic for the set of initial points lying on a periodic orbit: their participation
in the resolvent surface is restricted to a 2-dimensional torus, whatever the number of degrees of freedom,
that is, if $\x_-$ belongs to a periodic orbit, no point $\x_+$ outside this orbit can pair with it in the resolvent surface.

A further general property of the resolvent surface is that it must touch the identity plane 
at all periodic points $\x_+= \tilde{\x}(\x_-, k\tau)$, that is, even though a periodic orbit wanders off into double phase space,
it must return to the identity plane at multiples of its period $\tau$. Neighbouring points not initially on this periodic orbit
do not touch again the identity plane, unless their orbit is also periodic. In all cases, whatever the complexity of the sheets
of the resolvent surface, all its returns to the identity plane must lie on its energy shell $H(\x_+=\x_-=\x)=E$.
This special role of the identity plane gives the canonical centre and chord variables $\X=(\x, \Vxi)$ a privileged role 
in the description of the resolvent surface.

\section{Propagators, Green functions and Wigner functions}

Quantum operators form a Hilbert space of Hilbert-Schmidt operators 
with the {\it scalar product} \cite{Vor76,Littlejohn95},
\be
\langle\!\langle A|B\rangle\!\rangle= {\rm tr}\> \opA^{\dagger}\widehat B,
\label{scalarprod}
\ee
defined in terms of the adjoint operator, $\opA^{\dagger}$. Then, in this notation, the trace of an operator
is the scalar product with the identity operator $\opI$:
\be
{\rm tr} ~\opA = \langle\!\langle I|A\rangle\!\rangle ~.
\label{trace1}
\ee 

Each foliation of double phase space by parallel Lagrangian planes corresponds
to a possible operator representation. 
Perhaps the most common representation relies on the dyadic operators,
$\langle\!\langle \vQ|=|\vecq_-\rangle\langle \vecq_+|$, so that
\be
\langle\!\langle \vQ |A\rangle\!\rangle=\langle \vecq_+|\opA|\vecq_-\rangle=
{\rm tr}\>|\vecq_-\rangle\langle \vecq_+|\opA,
\ee
where the Lagrangian planes are just $\vQ=(\vecq_-,\vecq_+) =constant$. From this one can switch to momentum, 
or various mixed representations through Fourier transformations, corresponding to $90^o$
rotations in double phase space $\{\X=(\vP,\vQ)\} \mapsto \{\X'=(\vP',\vQ')\}$. 
As reviewed in Appendix A, the Wigner-Weyl representation, based on the self-adjoint operator,
$\opR_\x$, then corresponds to the double phase space rotation \eref{rotation}, so that for $\vQ'=\x$,
\be
\langle\!\langle \vQ'|A\rangle\!\rangle=2^L {\rm tr}\> \opR_{\x}\opA=A(\x),
\ee
and the Lagrangian basis in double phase space are the reflection planes, $\x=constant$.
The Fourier transformation \eref{FWS} then introduces the translation operator,
whose adjoint is $\opT_{-\Vxi}$. This is represented in double phase space 
by the new Lagrangian plane, $P'=\y=\J\Vxi=constant$, so that
\be
\langle\!\langle \vP'|A\rangle\!\rangle= {\rm tr}\> \opT_{-\Vxi}\opA=A(\Vxi).
\ee
In each case the representation in terms of a set of Lagrangian planes, $\vQ'$, is complementary
to the conjugate representation in terms of $\vP'$, which is obtained by a Fourier transform.
Of course, even though the trace \eref{trace1} remains invariant, it assumes different forms in each representation, for example
\begin{eqnarray}
{\rm tr} ~\opA &= \int {\rm d}\vQ~ \langle\!\langle I|\vQ\rangle\!\rangle \langle\!\langle \vQ |A\rangle\!\rangle
= \int {\rm d}\vecq_- {\rm d}\vecq_+ ~ \delta(\vecq_+ - \vecq_-) \langle\vecq_+|A|\vecq_-\rangle \\ \nonumber
&= \int {\rm d}\vecq \langle\vecq|A|\vecq\rangle ~~~~{\rm or} ~~~~ 
= \int {\rm d}\x~ A(\x) ~.
\end{eqnarray}
Further discussion is presented in  reference \cite{Chountasis}.

The evolution operator \eref{opevol} is represented by the various propagators $\langle\!\langle \vQ'|U_t\rangle\!\rangle$.
These are identified with their familiar notation in terms of ordinary phase space variables,
such as the position propagator $\langle\!\langle \vQ|U_t\rangle\!\rangle=\langle \vecq_+|U_t|\vecq_-\rangle$, or the Weyl propagator
$U_t(\x)$. In each case, the SC approximation assumes the standard form
\be
\langle\!\langle \vQ'|U_t\rangle\!\rangle \approx \sum_j A_{t,j}(\vQ') \exp \left[ \frac{i}{\hbar} \S_{t,j}(\vQ')+i \theta \right],
\label{Utdouble}
\ee
where the summation index $j$ labels all the conjugate points in the possibly many branches of the evolution surface $\vP'_{t,j}(\vQ')$;
for instance, all the pairs of momenta $\vecp_\pm$ for trajectory segments between $\vQ'=\vecq_\pm$, or all $\y= \J\Vxi$ for trajectory 
chords with tips centred on $\vQ'=\x$. Generally, the phase also includes Maslov indices \cite{Maslov} independent of $\hbar$, 
which will be denoted simply as $\theta$ in all the following formulae. 
The amplitude of each term depends on the stability matrix $\M_{t,j}$ for the relevant trajectory segment; 
e.g. the SC Weyl propagator is 
\be
U_t(\x) \approx \sum_j \frac{2^L}  {|\det (1+{\M_{t,j}(\x)})|^{1/2}} \exp \left[\frac{i}{\hbar}\S_{t,j}(\x)+i\theta\right] ~,
\label{SCUX}
\ee 
as reviewed in \cite{Alm98}.
On the other hand, the SC approximation for the trace of the evolution operator becomes \cite{AlmBro16}.
\begin{equation}
\langle\!\langle I|U_t\rangle\!\rangle = {\rm tr } ~ \widehat{U}(t) \approx \sum_j \frac{2^L}{|\det[1 - \M_{t,j}]|^{1/2}}\>\>
\exp \left[ \frac{i}{\hbar}S_{t,j}+i \theta\right] ~,
\label{Uweyl}
\end{equation}
where the sum is now over periodic orbits of period $t$ \cite{AlmBro16}. Their action $S_{t,j} =\S_{t,j}$, denoting the double space action over $\x_+$ on its own, does not depend on the particular representation in which it is derived.

Just as the Fourier transform in time of the evolution operator supplies the resolvent operator \eref{resolvent1}, 
the Fourier transform of the various representations of the evolution operators by propagators 
supplies the energy dependent Green functions, or spectral Wigner functions, which can be denoted in the general form
\be
\langle\!\langle \vQ'|U_E\rangle\!\rangle = \frac{1}{2\pi\hbar} \int_{-\infty}^{\infty} {\rm d}t  ~\exp \left[\frac{i}{\hbar} Et \right] ~
\langle\!\langle \vQ'|U_t\rangle\!\rangle ~.
\ee

Inserting the SC approximation for the propagators, the stationary phase evaluation of the time integral leads to 
a corresponding SC evaluation of the representation of the resolvent as
\be
\langle\!\langle \vQ'|U_E\rangle\!\rangle \approx \sum_j B_{EJ}(\vQ') \exp \left[ \frac{i}{\hbar} \S_{Ej}(\vQ') + i \theta \right] ~,
\label{UEdouble}
\ee 
in terms of the energy action \eref{enaction} for all $j$-trajectory segments on the $E$-shell with endpoints $\x_{\pm,j}(E)$, 
represented in Lagrangian coordinates as $\vP_j(\vQ',E)$.
This holds for all representations of the resolvent operator corresponding to conjugate Lagrangian planes in double phase space, which amount to different
perspectives of the same resolvent surface. Expressed locally by \eref{localsurf}, it does not have the open structure of
the evolution surface, so we first consider the simple case of a single degree of freedom in the following section.

As for the invariant trace of the resolvent operator, the SC approximation resulting from a stationary phase approximation 
of the Fourier transform has oscillatory terms of the form
\be
{\rm tr} ~ {\widehat U}_E = \langle\!\langle I|U_E\rangle\!\rangle \approx \sum_j C_{Ej} \exp \left[ \frac{i}{\hbar}\S_{Ej}+i \theta\right] ~,
\label{trUEdouble}
\ee 
summing over the po's in the $E$-shell, with their double action $\S_{Ej}$, which equals the ordinary po action $S_{EJ}$ in all representations. 
There is also a contribution from the limit of very short orbits, which varies smoothly with energy (e.g. reviewed in \cite{Ber91,Alm98}), but it will be left out of this presentation.

\section{Resolvent surface for one degree of freedom} 

In this case, the energy shell $H(\x)=E$, for a Hamiltonian describing bound motion, is a closed
curve $\gamma$ in the 2-dimensional phase space, topologically a circle. Then the trajectory
$\x_+= {\tilde \x}(\x_-,t)$ follows $\gamma$, closes it in the period $\tau(E)$
and retraces it infinitely many times. Therefore, there are infinitely many trajectory
segments for each pair of points on the curve: $\x_+= {\tilde \x}(\x_-,t+j\tau)$. 
These trajectory segments differ in action by $k$ times the periodic orbit (po) action
\be
\S_{E\rm po} =S_{E\rm po} =\oint_\gamma \vecp_+ \cdot {\rm d}\vecq_+.
\label{resaction}
\ee

Since the initial point $\x_-$ has been kept fixed, the circle in the double phase space traced by 
${\tilde \X}(t) = (\x_+= {\tilde \x}(\x_-,2t), \x_-)$ 
projects onto the $\x_+$-phase plane exactly as an ordinary po.
Clearly, one obtains the same projection for any initial point on the po, 
so that the full resolvent surface, that contains all pairs of points connected by a trajectory segment, 
is the product of the pair of circles in the $\x_\pm$ planes, i.e. a 2-dimensional torus.
If one considers the pair of curves in both the initial and the final phase spaces, their double action
cancels out according to \eref{canonical2}. Indeed, this full double phase space curve $\Gamma= (\gamma_-=\gamma,\gamma_+=\gamma)$
lies on the intersection of the resolvent surface with the evolution surface, which is always topologically a plane,
so all such circuits have zero double action as they wind around the torus that is the resolvent surface.  

Depending on the choice of Lagrangian coordinates, this torus is described by a function
$\vP'(\vQ')$, which is not univalued. Indeed, choosing the position representation $\vQ=(q_-,q_+)$
the projection onto this plane is a rectangle with four sheets, bounded by sides at the turning points 
where the $q_\pm$-components of the respective phase space velocities are zero, e.g. $V(q_\pm)=E$ for ordinary
Hamiltonians $H(\x) = p^2 /2m + V(q)$. This is also the projection in the momentum representation
$\vQ'=\vP=(-p_-, p_+)$, but there will be more sheets generated by extra internal folds
where $dV/dq=0$, if the potential does not increase monotonically, so that the energy shell
has a (classical) guitar shape, for instance.

Any symplectic ({\it linear canonical}) transformation in double phase space $(\vP,\vQ) \mapsto (\vP',\vQ')$ 
determines alternative views of the resolvent torus, in which generally the projection singularities separating the sheets 
will unfold into simple fold caustics between just a pair of sheets, as described in \cite{AlmHan82}. Remarkably,
the global singularity structure for the Wigner function, resulting from the transformation $(\vP,\vQ) \mapsto (\x, \Vxi)$ in \eref{rotation},
represents the resolvent surface $\Vxi(\x)$ with only two sheets for $\x$ close to the energy shell. 
As noted before, the points $\x$ are centres of chords, which have a double role: as true geometrical chords 
between a pair of points in the ordinary phase space, together with that of the double phase space variable $\y=\J\Vxi$.
It is easy to see that there are two geometric chords for each centre by merely exchanging their endpoints $\x_\pm$.
These coincide only for $\y=\J\Vxi=0$ on the shell at the identity plane itself where $\x_+ =\tilde{\x}(\x_-,k\tau)$, 
with $\tau$ the period of the orbit. 

In these double phase space coordinates the energy surface within the centre or identity plane is a projection singularity of the resolvent surface, 
since $\x_E(\Vxi)$ is an even function and $\Vxi=0$ within this plane, according to \eref{rotation}. 
Thus, the pair of chords centred on $\x$ coalesce as this point approaches the shell from its interior.
Starting from any point $\x(0)=\x_+=\x_-$ on the shell, the centre $\x(t)$ follows $\x_+(t)=\tilde{\x}(\x_-,t)$ to the interior of the shell, also describing a closed curve 
as the period is completed, 
though the chord $\Vxi(\x,t)=\tilde{\x}(\x_-,t)-\x_-$ reverses its sign.   
For $\x$ far inside the shell, it  becomes the centre of large chords stretching across this orbit and meets generically  
a cusped caustic island with three sheets in its interior. This configuration shown in Figure 2 was not described in \cite{AlmHan82}, 
but is present in Berry's original essay on SC Wigner functions \cite{Ber77}. 

\begin{figure}
\centering
\includegraphics[width=0.5\linewidth]{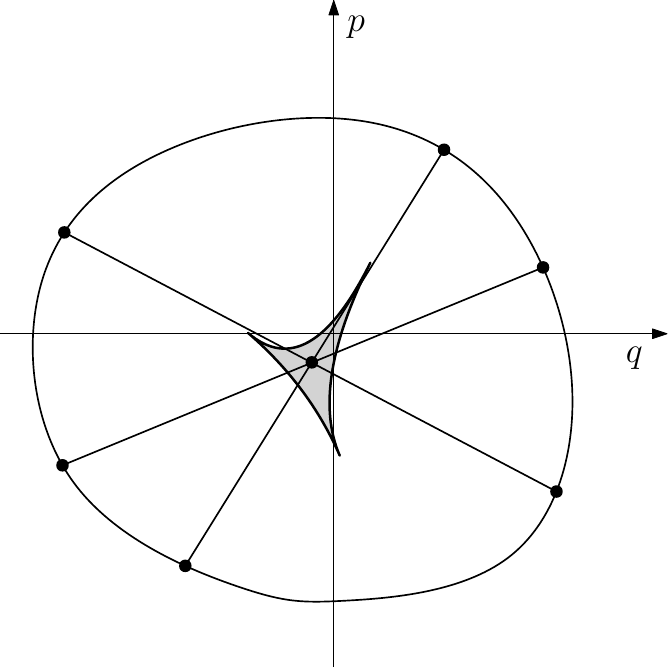}
\caption{A general convex closed energy shell in a 2-D phase space is also a periodic orbit and a caustic, i.e. a projection singularity of the 2-D resolvent surface onto the centre plane: it bounds the inner region, where each centre has a pair of chords $\pm\Vxi(\x)$ from the outer region with no chords. Deep in the interior, another caustic shaped as a cusped triangle
bounds a region where there are three pairs of chords for each centre $\x$.
\label{Fig2}}
\end{figure}

Even though the resolvent torus has not been previously mentioned in the literature, its projection singularities 
are precisely the caustics where the SC approximation for various representations diverge, that is, an amplitude
$B_{E,j}(\vQ') \rightarrow \infty$ in the sum \eref{UEdouble}. Strictly, these SC approximations
concern representations of the resolvent, but the same formulae may be extended to a projector $|n\rangle \langle n|$,
i. e. the residue of a pole in \eref{resolvent1}. Heuristically, it is evident that only for an energy shell quantized
by the SC Bohr-Sommerfeld rules can the sum over the infinite repetitions of the trajectory segments add up in phase
to produce the required pole. This argument can be further developed, though the final result, based simply on single segments
without repetitions, are more easily obtained in other ways. In short, the SC caustic singularities of the representations
of the resolvent for energies between its poles, as well as the residues at the poles, i.e. the projectors,
lie on the singularities of the projections of the resolvent surface onto each chosen Lagrangian coordinate system.

\section{The resolvent surface as a chaotic sponge}

Let us now proceed directly to a nonintegrable system with $L>1$ degrees of freedom, since the integrable case
treated in Appendix B presents a completely different generalization of the previous section. 
The trajectories in ordinary phase space for this more general scenario are not constrained to $L$-dimensional tori and, even if such exist,
they are generally nonresonant, that is, they are not families of periodic orbits,
which, in their turn, will be presumed to be isolated in each energy shell. 

The conservation of energy along each trajectory $\x_+={\tilde \x}(\x_-,t)$ implies that the resolvent surface is the extension of the energy shell by all the trajectories that depart from the $L$-dimensional identity plane
$H(\x_+=\x_-=\x)=E$, in the course of positive and negative time. As previously noted,
for short times the resolvent surface grows cylindrically according to  \eref{initialcyl} 
in the directions of the double phase space velocities generated by the double Hamiltonian \eref{Scham}.
For slightly longer positive times the trajectories bend inwards, if the shell is locally convex,
with their (geometric) chords centred inside the shell like the string of a bow. Therefore, the shell on the identity plane 
is a simple fold caustic of the resolvent surface as viewed from its centre variables $\x$. Regardless of the number of degrees of freedom,
this short time behaviour is entirely analogous to that discussed in the previous section,

However, this simple structure of the growing cylinder will only last until $|t|=\tau_1$, the shortest period of a 
periodic orbit. Then, by the same argument as in the case of a single degree of freedom, which
holds whatever the dimension of the phase spaces, each point in this first po generates a circle
and this continuum of circles forms a 2-dimensional torus. The phase space action \eref{resaction}
has the same form as that for a single degree of freedom. 
Indeed, as it lies in a Lagrangian surface, 
the double phase space action, including both the $\x_-$-circuit and the $\x_+$-circuit, cancels again.
Still, one must keep in mind that this two-dimensional torus is now embedded in the $2L$-dimensional resolvent surface. Thus there is no inside or outside
of this torus and neighbouring trajectory segments almost retrace themselves.

The essential novelty imparted by the increased dimension is that each repetition of the po is connected to an extension 
of the resolvent surface
in its neighbourhood. Indeed, the smoothness of this surface generated by a continuum of smooth trajectories demands that neighbouring trajectories to the po also nearly touch the identity plane at its period. Furthermore,
it is deduced in \cite{Alm98} that a fold caustic of the resolvent surface projected on the centre plane, 
touches the po, which itself lies on the caustic along the entire energy shell, as sketched in Figure 3. 
The initially narrow po caustic parts from the energy shell and broadens as it recedes into the interior.
This implies that, added to the short chords of centres within this narrow fold, 
there are chords of pairs of neighbouring trajectory segments to the po, 
which do not quite retrace themselves, so that they only approach the centre plane without touching it again. 
Thus, one might say that a pair of thin tongues develop in the resolvent surface itself, 
which reach down and lick the the initial periodic orbit in the centre plane precisely at $t =\pm\tau_1$.
Symmetry of the trajectory growth for positive and negative times allows us to concentrate on positive times hereon.  
Of course, this construction of the tongue of the resolvent surface at a particular point on the po can be extended 
all the way around the po, so that a circular tongue is formed.

\begin{figure}
\centering
\includegraphics[width=0.5\linewidth]{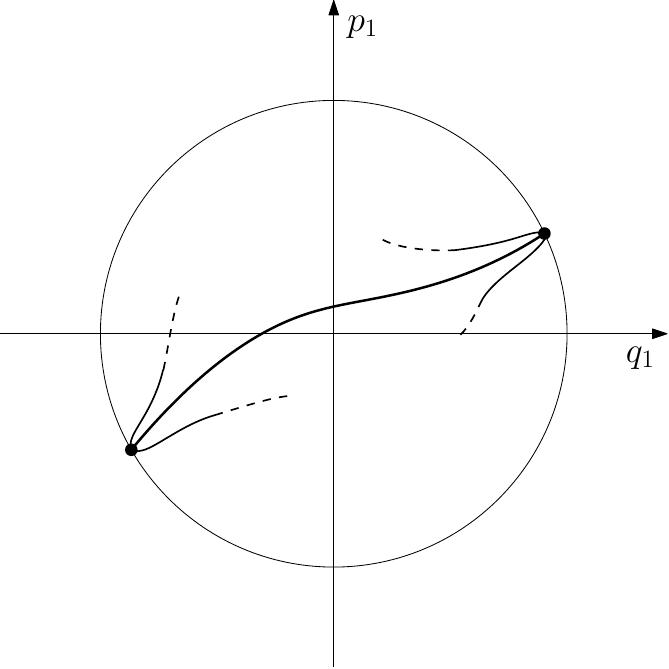}
\caption{The closed curve represents a section of the 3-D energy shell within the centre plane, a caustic, i.e. a projection singularity of the resolvent surface. The segment joining two points on this curve would be the projection of a periodic orbit. Whereas its internal structure is analogous to the previous figure, the curves sketched near its tips represent the surrounding caustics, the projection singularities of tongues of the resolvent surface. 
\label{Fig3}}
\end{figure}

That a new tongue is produced at each new period of the po is a consequence of the deduction  in \cite{Alm98} of its centre caustic.
\footnote{All trajectory segments contributing to the Weyl propagator for each centre $\x$ have their tips lying in its particular Poincar\'e section. This is a small glancing section 
if this point is close to the energy shell, so that orbits that enter the section from the outside 
can coalesce at a caustic according to \cite{Alm98}.} 
Even though the question of multiple windings was not treated explicitly there, 
the linearized action difference $S(\x)-S(\x_{po})$ depends on the stability matrix for the Poincar\'e  map of the po, 
which has a power proportional to the number of windings. 
Therefore, there is a different action and hence a different chord for each winding 
and thus a new lip of the resolvent surface is formed at each return to the po.  

In this way the analysis of the caustics in \cite{Alm98} supplies direct information about the growth of the resolvent surface with time.
A first general picture is then that all trajectories depart vertically
from the energy shell at the centre plane. This includes each periodic orbit, which has no distinction
before its first period. It is only as this period is reached that it touches again the initial po
on the centre plane, bringing along with it a narrow tongue of the resolvent surface. Then this process repeats
itself, bringing a new tongue to the initial orbit at each iteration... However, this is only the first po to complete its cycle
and the growth of narrow tongues of the resolvent surface that reach down to the centre plane, 
as the period of each successive po is reached, successively enriches its structure. It is an open question how these tongues
broaden and blend into smooth folds of the resolvent surface over each centre $\x$ deep inside the the shell on the centre plane. 
Nonetheless, the resultant multiple layered structure may justly be described as a chaotic sponge!

\section{Secondary periodic orbits}

For a chaotic system, the isolated po's in each $(2L-1)$-dimensional energy shell are mostly hyperbolic.
Then there are $(L-1)$ directions of nearby points in the energy shell that
approach the po asymptotically in time and $(L-1)$ directions for which this approach occurs for $t\rightarrow - \infty$.
These directions are tangent to the $L$-dimensional stable and the unstable manifolds respectively, which are known to play an important role in SC approximations \cite{LiTom}. Here, one just points out that a local extension of the growing resolvent surface for all positive and negative times
joins each double phase space po, for which $\x_+= \tilde{\x}(\x_-,\tau_j)=\x_-$, to its stable and unstable manifolds.

One should note that the full extension of the resolvent surface for very long positive and negative times
generates a horribly convoluted $(2L)$- dimensional surface within the $2(2L-1)$-dimensional direct product
of the pair of energy shells $H(\x_\pm)=E$. Generally, one need only consider times up to the order of the
Heisenberg time $t_H {\approx} \hbar /\overline{(E_n-E_{n-1})}$, but this is insufficient to connect
a po to its stable and unstable manifolds. Therefore, one needs here to place an initial point $\x_-$
on this manifold and propagate it for infinite time, so that it reaches its nearby po. 
Further extensions of the resolvent surface beyond the Heisenberg time will not be required in this essay.    

The $L$-dimensional unstable manifold develops along $L-1$ the unstable directions transverse to each point of the po, 
thus forming an $L$-dimensional cylinder. This topology is preserved, despite eventually wrinkling in a very complex manner. 
Being Lagrangian in the original phase space, the action around any irreducible circuit on this cylinder
equals that of its parent po  \cite{OzMehtaClay} and an analogous picture holds for the stable manifold in reverse time. 
The $L$-dimensional stable and unstable manifolds are allowed to intersect in the $(2L-1)$-dimensional energy shell, 
so defining a homoclinic orbit, which is bi-asymptotic to the parent po for $t\rightarrow \infty$.
Likewise, an intersection of the unstable manifold of po$_j$ with the stable manifold of po$_{j'}$
creates a heteroclinic orbit, originating in the neighbourhood of po$_j$ and then winding infinitely
around po$_{j'}$. 

Consider now a path that winds $k(j)$ times around the parent $j$-po, then makes a small step (across other trajectories) 
to the homoclinic orbit and then another step, so as to close itself at the initial point. 
This is not a trajectory, but a loose invocation of the
shadowing lemma \cite{Palmer} indicates that there should exist a true po that lies very close to this path 
and, hence, winds $k(j)$ times around the $j$-po.
Indeed, the algorithm developed by Michel Baranger to converge on periodic orbits \cite{Bar} is applicable to the accurate computation
of such secondary periodic orbits in Hamiltonian systems \cite{Vieira96}, as well as in canonical maps \cite{RitAlmDou}.

Analogous steps can be followed for a departure from the $j$-po, after winding around it $k(j)$ times, along a heteroclinc orbit
and then winding $k'(j')$ times around the $j'$-po; but then one needs to return along
a complementary heteroclinic orbit back to $j$-po, so as to close the circuit. This serves as a blueprint
for a sequence of secondary periodic orbits that accumulate on the pair of heteroclinic orbits as the windings $k(j)$ and $k'(j')$ are increased.
Indeed, there is no need to stop at two primitive po's:
any number of them are connected sequentially by the many intersections of their stable and unstable manifolds.
Infinite sequences of secondary po's, with higher and higher windings along each of these primitive po's,
accumulate along the heteroclinic orbits in each of these constructions and it is observed computationally that the secondary po's
become almost indistinguishable from the parent po's even for moderately low windings $k(j)$ \cite{RitAlmDou}.

Even though the secondary po's are constructed as windings around the primitive po's together with their heteroclinic excursions,
they are true periodic orbits. Therefore, as each of their large periods is completed, a thin tongue will descend in the double phase space
onto this newly completed secondary po on the identity plane, very close to the tongues, which lick the component primitive po's.
Thus there will be many fold caustics in the vicinity of a short po, which delimit regions where longer and longer trajectory segments
nearly return to their initial point.

The emphasis on closed secondary periodic orbits should not obscure the presence of similar constructions that decorate open orbits.
One shall assume that the set of homoclinic orbits asymptotic to a single po, or the set of heteroclinic orbits is just as dense 
among the full set of trajectories as the the set of po's themselves. 
\footnote{It may seem optimistic for a simple open orbit to wander close to a homoclinic or heteroclinic orbit. 
However, in the case of a Bernoulli system \cite{ArnAv}, in which the chaotic orbits are described in terms of a finite alphabet, it is easy to see
that the set of homoclinic orbits to a single po is dense, among the full set of orbits.}
Therefore any short open orbit will come close to a homoclinic
or a heteroclinic orbit of primitive po's and, hence, any number of cycles around primitive po's can be added to construct 
long open orbits, which satisfy the same boundary conditions, such as a pair of end-positions $\pm\vecq$, or a fixed centre $\x$ between the endpoints $\pm\x$.

\section{Secondary orbits and composite orbits in the semiclassical scenario}
	
A general secondary po is characterized by its primitive po's, $j1,, j2, ...$, which will be represented by the capital $J$,
and the windings around each primitive po, $k1(j1), k2(j2), ...$, labled simply as $K(J)$. In the limit 
where all the $kn(jn) \rightarrow \infty$, expressed as $K(J) \rightarrow \infty$, the windings around each primitive po
accumulate on it and their connection to the windings around other po's approach the heteroclinic orbits. For short,
one may define $\S_{EK(J)\rm het}$ as the action of the total heteroclinic circuit closed among
the primitive po's. (Indeed, there are many heteroclinic orbits joining a given pair of po's, ordered by their growing length. 
Each of these belongs to a different blueprint for a family of secondary periodic orbits). 
Thus, one obtains the limit of the total
action of the secondary po as 	
\be
\S_{EK(J)} \rightarrow  \S_{EK(J)\rm het}+\sum_{j,k(j)}  k(j) \S_{Ej}~,
\label{actsec}
\ee	
where it is of note that the second term, a linear combination of the actions of a set of primitive po's,
coincides with the action of a pseudo orbit \cite{Ber-Keat90,Ber-Keat92}, or composite orbit \cite{Sieber07} 
\be
\S_{EK(J)\rm com} = \sum_{j,k(j)}  k(j) \S_{Ej} ~,
\ee
which made its debut in the theory of quantum spectral determinants for individual states of chaotic systems.
 
A similar form describes the action of an open trajectory, which wraps many times around a circuit of primitive po's,
neighbouring a secondary po, on its way between two given endpoints. Somewhere along its way, a short open orbit 
will approach a heteroclinic orbit of a circuit, allowing for a connection to it across a continuous set of trajectories. 
Then in the limit $K(J) \rightarrow \infty$ the action for this 'dressed' open orbit acquires the 
total action of the $J$-circuit, just as the secondary $K(J)$-po, so that the action of this secondary open orbit becomes
\be
\S_{EK(J)}(\vQ) \rightarrow \S_{E,0}(\vQ)  + \S_{EK(J)},
\label{actopen2}
\ee
for any primitive open orbit with action $\S_{E,0}(\vQ)$, with the appropriate endpoints represented by the double phase space variable $\vQ$.

Therefore, the SC contribution of the simple primitive po's to the resolvent and its trace for long times, beyond the Heisenberg time,
may be approximately grouped within the composite orbits.  
Starting with the oscillatory part of the trace of the resolvent, we can insert \eref{actsec} into \eref{trUEdouble} 
to produce a sum over the long time (LT) secondary orbits labled by $J$ 
\begin{eqnarray}
\{{\rm tr} ~ {\widehat U}_{EK(J)}\}_{\rm LT} \approx \sum_{K(J)} C_{EK(J)} \exp \left[ \frac{i}{\hbar}\S_{EK(J)}+i \theta\right]  ~.
\label{trUEdouble2}
\end{eqnarray}
(Note that the special case of a single primitive orbit is here included as a secondary orbit with only one nonzero $j$-entry.)
The long time contribution of the short primitive orbits may then be expressed as 
\be
\fl \{{\rm tr} ~ {\widehat U}_{EK(J)}\}_{\rm LT} \approx \sum_{K(J)} C_{EK(J)}\exp \left[ \frac{i}{\hbar} \S_{{\rm het}EK(J)} \right] 
~\exp \left[ \frac{i}{\hbar}\sum_{j,k(j)} k(j) \S_{Ej}+i \theta\right]~,
\ee
recalling that the amplitude of each secondary po receives a complex phase from its heteroclinic excursions. 
From this point of view, the action of all the composite orbits (or pseudo-orbits) with a long total period
are already present in the trace of the resolvent. 
The resummation of the density of states consummated by spectral determinants \cite{Ber-Keat90,Ber-Keat92,Bog92,Sieber07}
can then be considered to shift the contribution of the compound orbits, already present in the trace, 
from long total periods to those that are shorter than the Heisenbrg time.

Still, this is a bona fide periodic orbit,
so that it brings down a new tongue of the resolvent surface as its period is completed, thus forming a new layer of the chaotic sponge.  
In the long time limit the amplitude of the contribution to the trace of each secondary orbit may also be derived from its primary components.
Indeed, its main ingredient, already present in the amplitude of the trace of the evolution operator \eref{Uweyl}, 
is the stability matrix for the entire orbit. This is dominated by the product of primitive stability matrices
\be
\M_{t,K(J)} \approx \prod_{j, k(j)} \M_{t,j}^{k(j)}, 
\ee
if one neglects the short heteroclinic excursions. (This decomposition does not apply to a secondary po with few primary windings). 

A similar decomposition holds for the contributions to the representations of the resolvent of short open orbits, 
which can be dressed by a secondary po. Labling the dressed open orbit by $K(J)$, i.e. by the added secondary orbit, 
one obtains its contribution by inserting \eref{actopen2} in \eref{UEdouble}
\begin{eqnarray}
\{\langle\!\langle \vQ'|U_{EK(J}\rangle\!\rangle\}_{\rm LT} \approx  B_{EK(J)}(\vQ') \exp \left[ \frac{i}{\hbar} \S_{EK(J)}(\vQ') + i \theta \right]  \\  \nonumber
=B_{EK(J)}(\vQ') \exp \left[ \frac{i}{\hbar} \S_{E0}(\vQ') \right] ~\exp \left[ \frac{i}{\hbar} \sum_{j,k(j)} k(j)\S_{Ej}(\vQ') + i \theta \right] ~.
\label{UEdouble2}
\end{eqnarray}
Hence, again, the resummation of the Green function or the Wigner function basically exchanges the contribution 
of long open composite or pseudo orbits by those with a period shorter than the Heisenberg time.

\section{Discussion}

The extraordinary complexity of the flow generated by an unintegrable Hamitonian was first put forward by the startling presentation
of homoclinic intersections by Poincar\'e \cite{Poincare}. Even though he did not emphasize this aspect, each unstable (stable)
$L$-dimensional manifold emanates from each periodic orbit (po) as an initially well behaved cylinder, formed by helical trajectories.
Further on, the cylinders become evermore wrinkled, so that their folds eventually explore the entire energy shell. Nonetheless,
they never loose their smoothness, just as their constituent trajectories, and any circuit around them
reproduces the action of the parent po \cite{OzMehtaClay}. Their one-dimensional intersections with other stable (unstable) manifolds
define the homoclinic or heteroclinic orbits, 
which approach bi-asymptotically their respective po's 
as they wrap helically around the stable (unstable) cylinder.
In this way, one achieves a fairly simple description of this exotic creature, the heteroclinic orbit,  
by sharing attention between both the initial unstable cylinder and the final stable cylinder.

There is some parallel of this microscopic picture of Hamiltonian chaotic motion to its macroscopic description in terms of the 
resolvent surface. Instead of picking a single isolated po, it is the entire energy shell, within the $2L$-dimensional identity plane
in the $4L$-dimensional double phase space, that seeds a cylinder as each trajectory grows from its initial point. In contrast with
the evolution of an $L$-dimensional unstable manifold from a simple circuit, a topological circle, one just adds one more dimension 
to the $(2L-1)$-dimensional shell in the doubled phase space. Notwithstanding its trivial projections 
onto the initial $\x_-$-plane and the final $\x_+$-plane, which are identical with the energy shell itself, 
the full resolvent surface develops smooth folds in the form of narrow tongues, which reach back onto the identity plane along each po.
The enormous multiplicity of smooth narrow tongues, reaching back onto all the primary and secondary po's in the identity plane, admits the description of the resolvent surface
as a multidimensional sponge, even if one cuts off the period of all the trajectories at the Heisenberg time. 

Such is the complex geometrical structure that supports semiclassical approximations of various representations of the resolvent operator.
They differ only by the choice of Lagrangian coordinate planes for the description of the resolvent surface, which is itself Lagrangian
in the double phase space. In each of the chosen coordinates, the phase of the semiclassical contribution of an orbit segment is an integral along the resolvent surface, in strict analogy to the phase of a semiclassical wave function in ordinary $2L$-dimensional phase space.
The existence of multiple folds generated by the po's is here deduced from the caustics within the conjugate Wigner-Weyl coordinates 
of centre and chord, but it is not conceivable that projections of such an intricate structure onto other double phase space coordinates
can avoid multiple projection singularities. In the neighbourhood of each of these caustics, the simple semiclassical approximation for
the corresponding representation of the resolvent operator breaks down and must be replaced by a more refined uniform approximation \cite{BerUp}.

The fact that periodic orbits are dense within each energy shell may be claimed for the plausibility of their role in the semiclassical
approximation of the quantum density of states, that is, the trace of the resolvent. However, the multiple families of secondary po's,
that are highlighted in this essay, wind evermore tightly around a small set of primitive po's, even if their heteroclinic connections
wander somewhat freely between them. The distinction between primary and secondary periodic orbits is robust with respect to any softening 
of the chaos, which allows for the presence of small islands surrounding stable periodic orbits. They will form two-dimensional tori 
within the resolvent surface (just as in the integrable systems described in Appendix B), which in no way impede the heteroclinic connections
and the formation of secondary po's in their neighbourhood.   

If one accepts that these high period secondary po's are resummed 
in spectral determinants into composite orbits (with actions that are linear combinations of the actions of their primitive constituent po's),
then it appears that it is a relatively small set of primitive po's that really matter as the classical skeleton of quantum mechanics.
An indication that this conjecture may be correct is that, in the very simple example of the spectral determinant 
for an integrable map (with no secondary po's) \cite{OzTomLew}, resummation of the density of states produced inferior results to the straight semiclassical trace. It is notable that a recent alternative scheme for building quantum states around a small set of hyperbolic po's 
showed promising computational results \cite{Revetal}.

 \appendix

\section{Quick review of the Wigner-Weyl (centre) representation}

Following the notation of the review \cite{Alm98},
let us recall that each chord coordinate in double phase space $\Vxi=-\J\y$ lables a uniform translation of phase space
points $\x_- \in \bf{R}^{2L}$ by the vector $\Vxi \in \bf{R}^{2L}$,
that is: $\x_- \mapsto \x_- + \Vxi$. Likewise, each centre, $\x$,
labels a reflection of phase space through the point $\x$, 
that is, $\x_- \mapsto 2\x - \x_-$.

Corresponding to the classical translations, one defines {\it translation operators},
\begin{equation}
\hat{T}_{\Vxi} = \exp\left\{\frac{i}{\hbar}\,\Vxi\wedge \hat{\x}\right\} ,
\end{equation}
also known as displacement operators, or Heisenberg operators.
The chord representation of an operator $\hat{A}$ on the Hilbert
space is defined as the decomposition of
$\hat{A}$ into a linear (continuous) superposition of translation operators.
In this way,
\begin{equation}
\label{conC} 
\hat{A} = \frac{1}{(2\pi\hbar)^L}\int d\Vxi \;
\tilde{A}(\Vxi)\; \hat{T}_{\Vxi} \
\end{equation}
and the expansion coefficient, a function on $\bf{R}^{2L}$, is the
{\it chord symbol} of the operator $\hat{A}$:
\begin{equation}
\label{covC} 
\tilde{A}(\Vxi) = {\rm tr} \;\left[\hat{T}_{-\Vxi}\;\hat{A}\right]  .
\end{equation}

The Fourier transform of the translation operators defines the
{\it reflection operators},
\begin{equation}
2^N\;\hat{R}_\x = \frac{1}{(2\pi\hbar)^L}\int d\Vxi \;
\exp\left\{\frac{i}{\hbar}\,\x\wedge\Vxi\right\} \; \hat{T}_{\Vxi}  ,
\label{refl}
\end{equation}
such that each of these corresponds classically to a reflection of phase space
$\bf{R}^{2L}$ through the point $\x$.
The same operator $\hat{A}$ can then be decomposed into a
linear superposition of reflection operators
\begin{equation}
\label{conW} 
\hat{A} = 2^L\int \frac{d\x}{(2\pi\hbar)^L}\; A(\x) \;
\hat{R}_{\x} ,
\end{equation}
thus defining the {\it centre symbol or Weyl symbol} of operator $\hat{A}$,
\begin{equation}
\label{covW} 
A(\x) = 2^L{\rm tr}\;\left[\hat{R}_{\x}\;\hat{A}\right]
\end{equation}
In the case of the density operator $\hat \rho$ (the residue of the resolvent operator for a pure state), it is convenient to use another normalization for 
the {\it Wigner function} \cite{Wigner}:
\begin{equation}
W(\x) = \frac{1}{(\pi\hbar)^L}{\rm tr}\;\left[\hat{R}_{\x}\;\hat{\rho}\right].
\label{Wtr}
\end{equation} 

The centre and chord symbols are always related
by a Fourier transform:
\begin{equation}
 \tilde{A}(\Vxi) = \frac{1}{(2\pi\hbar)^L}\int d\x 
 \exp\left\{\frac{i}{\hbar}\,\x\wedge \Vxi\right\}
A(\x)
 \ ,
\label{FWS}
\end{equation}
\begin{equation}
 A(\x) = \frac{1}{(2\pi\hbar)^L}\int d\Vxi\; 
\exp\left\{\frac{i}{\hbar}\,\Vxi\wedge \x\right\} 
\tilde{A}(\Vxi).
\label{FCS}
\end{equation}
In particular, one obtains the reciprocal representations of the reflection operator and 
the translation operator as
\begin{equation} 
2^L\;\tilde{R}_{\x}(\Vxi)= \exp \left\{\frac{i}\hbar \,\x \wedge \Vxi\right\}\;\;\; {\rm or}\;\;\;
T_{\Vxi}(\x) = \exp \left\{-{\frac{i}\hbar}\, \x \wedge \Vxi\right\}. 
\label{planewave} 
\end{equation}
These expressions are ideally suited for use in SC approximations,
because they are already expressed exactly by a classical phase: a plane wave in double phase space. The direct representations are 
\begin{equation} 
2^L\;\tilde{R}_{\x}(\x')= \delta(\x'-\x)\;\;\; {\rm or}\;\;\;
T_{\Vxi}(\Vxi') = \delta(\Vxi'-\Vxi).  
\label{directrep}
\end{equation}

\section{Integrable resolvent surface}

The traditional SC theory for integrable systems deals directly with individual eigenstates and so dispenses the bipass through the resolvent operator and its surface, that is, it relies on $L$-dimensional Lagrangian surfaces. The trajectory segment between pairs of points is then not needed and one can even deal with dyadic operators between pairs of states  $|n\rangle \langle n'|$ \cite{Dod86}. Nevertheless, it is worth describing the integrable extension of the resolvent surface for completeness.   

According to Arnold's theorem \cite{Arnold78} the trajectories of an integrable classical system of $L$ degrees of freedom
are confined to $L$-dimensional tori within the $(2L-1)$-dimensional energy shell. At first sight, one could try to construct
the resolvent surface by taking again the product of the tori in the initial space $\x_-$ with their identical image in $\x_+$,
but then each such double torus already has $2L$ dimensions, i.e. the same as the entire resolvent surface. In contrast,
the product of the pair of energy shells $H(\x_\pm)= E$ has $2(2L-1)$ dimensions, which are too many for a Lagrangian surface
in the  $4L$-dimensional double phase space, even though the resolvent surface must lie within this higher dimensional surface.

To clarify the predicament, let us restrict $L=2$ and consider a resonant torus, i.e. a continuous loop of periodic orbits.
Now the reasoning in section 5 applies for each individual po, which then generates a 2-dimensional po-torus.
Clearly, following a continuous sequence of loops around the original torus, one then generates a 3-dimensional torus, contained within the
4-dimensional product torus. But, of course, the product torus contains all pairs of points
in the original torus, whereas the resolvent surface selects only those pairs connected by a trajectory and so excludes
pairs of points lying on different po's.  
If all the tori are resonant, as with the double harmonic oscillator with comensurate frequencies,
then the 4-dimensional resolvent surface is foliated by these resonant 3-dimensional tori. Even for a nonresonant
system, the resonant tori are dense, so that the contribution to the resolvent surface from a nonresonant surface
squeezed between them must also be 3-dimensional.

A direct construction of such a 3-dimensional leaf of the resolvent surface in double phase space follows the procedure in section 6
of turning on the time and following the trajectories starting off on the initial torus. This generates a time dependent
mapping of the original 2-dimensional torus to itself, which in the action-angle variables $\x=(I_1,I_2,\theta_1,\theta_2)$,
appropriate for integrable systems \cite{Arnold78}, is merely the translation (within a torus with fixed actions $(I_1,I_2)$)
\be
\theta_{1+}=\theta_{1-}+\omega_1(I_1,I_2) t  ~~~ {\rm and} ~~~~  \theta_{2+}=\theta_{2-}+\omega_2(I_1,I_2) t ~,
\ee
where each frequency
\be
\omega_m(I_1,I_2) = \frac{\der H}{\der I_m}(I_1,I_2) ~.
\ee
The 3-dimensional surface is then the locus of these pairs of points $\x_\pm(t)$ for for all time. Thus, even though
the nonresonant leafs of the resolvent surface wind on forever, they are narrowly confined within the product shell, 
while approaching arbitrarily close to any point on the product torus. In any case, one perceives that, 
even with complete integrability, the decomposition of the resolvent surface into tightly intermingling closed 
and open surfaces of lower dimension is much more complex than the simple picture for a single degree of freedom.

\ack
I thank Gabriel Lando and Raul Vallejos for stimulating discusssions.
Partial financial support from the
National Institute for Science and Technology--Quantum Information
and CNPq is gratefully acknowledged.

\end{document}